\documentclass[aps,prl,twocolumn,showpacs,amsmath,amssymb]{revtex4}

\usepackage{graphicx}
\usepackage{dcolumn}
\usepackage{bm}

\newcommand{\ket}[1]{\ensuremath{\vert{#1}\rangle}}

\begin{document}

\preprint{Preprint}

\title{Conductance in strongly correlated 1D systems: 
Real-Time Dynamics in DMRG}

\author{Guenter Schneider}
\email{gs@tkm.uni-karlsruhe.de}
\author{Peter Schmitteckert}
\email{peter@tkm.uni-karlsruhe.de}
\affiliation{%
Institut f\"ur Theorie der Kondensierten Materie, Universit\"at
Karlsruhe, 76128 Karlsruhe, Germany
}%

\date{\today}

\begin{abstract}
A new method to perform linear and finite bias conductance
calculations in one dimensional systems based on the calculation of
real time evolution within the Density Matrix Renormalization Group
(DMRG) is presented.
We consider a system of spinless fermions consisting
of an extended interacting nanostructure attached to
non-interacting leads. 
Results for the linear and finite bias conductance through a
seven site structure with weak and strong nearest-neighbor
interactions are presented. Comparison with exact diagonalization
results in the non-interacting limit serve as verification of the
accuracy of our approach. Our results show that interaction effects
lead to an energy dependent self energy in the differential
conductance.
\end{abstract}



\maketitle


During the past decade improved experimental techniques have made
production of and measurements on one-dimensional systems possible
\cite{Sohn_Kouwenhoven_Schon:1997}, and hence led to an increasing
theoretical interest in these systems. 
However, the description of non-equilibrium transport properties,
like the finite bias conductance of an interacting nanostructure 
attached to leads, is a challenging task.
For non-interacting particles, the conductance can be extracted from the
transmission of the  single particle levels \cite{Landauer57,Landauer70,Buettiker86}.
Since the screening of electrons is reduced by reducing the size of structures
under investigation, electron-electron correlations can no longer be neglected.
Recently several methods to calculate the zero bias conductance of strongly interacting 
nanostructures have been developed. One class of approaches consists in extracting the conductance
from an easier to calculate equilibrium quantity,
e.g.\ the conductance can be extracted from a persistent current calculation
\cite{Molina_Schmitteckert_Weinmann_Jalabert_Ingold_Pichard:2004},
from phase shifts in NRG calculations \cite{Oguri05},
or from approximative schemes based on the tunneling density of states \cite{MeirWingreenLee91}.
Alternatively one can evaluate the Kubo formula within Monte-Carlo 
simulations~\cite{Louis_Gros:2003}, 
or from DMRG calculations~\cite{dan06}.
In contrast, there are no general methods available to get rigorous results
for the finite bias conductance. While the problem has been formally solved by
Meir and Wingreen using Keldysh Greens functions~\cite{MeirWingreen92},
the evaluation of these formulas for interacting systems is generally based on approximative schemes.

In this work we propose a new concept of calculating finite bias conductance
of nanostructures based on real time simulations
within the framework of the DMRG
\cite{White92,White93,cazalilla02,luo03,daley04,white04,peter04,white05}.
It provides a unified description of strong and weak
interactions and works in the linear and finite bias regime, as long as
finite size effects are treated properly.

In a first approach of real time dynamics within DMRG, Cazalilla and Marston integrated
the time-dependent Schr\"odinger equation in the Hilbert space
obtained in a finite lattice ground state DMRG calculation~\cite{cazalilla02}.
Since this approach does not include the density matrix for the time evolved
states, its applicability is very limited.
Luo, Xiang and Wang~\cite{luo03} improved the method by 
extending the density matrix with the contributions of the wave function
at intermediate time steps.
Schmitteckert~\cite{peter04} showed that the calculations 
can be considerably improved by replacing the integration of
the time dependent Schr{\"o}dinger equation with the evaluation of
the time evolution operator using a  Krylov subspace method for matrix
exponentials and by using the full finite lattice algorithm.

An alternative approach is based on the wave function prediction
\cite{WhitePredicition}. There one first calculates an initial state
with a static DMRG. One iteratively evolves this state
by combining the wave function prediction with a time evolution scheme.
In contrast to the above mentioned full t-DMRG, one keeps only
the wave functions at two time steps in
each DMRG step.
In current implementations the time evolution
is calculated by approximative schemes, like the Trotter
decomposition~\cite{daley04,white04}, or the Runge-Kutta method~\cite{white05}.
In our work, we combined the idea of the adaptive DMRG method with
direct evaluation of the time evolution operator via a matrix
exponential as described in Ref.~\cite{peter04}. 
Therefore our method involves no Trotter approximations,
the time evolution is unitary by construction, and
it can be applied to models beyond nearest-neighbor hopping.

\begin{figure}[tb]
\begin{center}
\newlength{\graphiclength}
\setlength{\graphiclength}{0.4\textwidth}
\includegraphics[width=\graphiclength,clip]{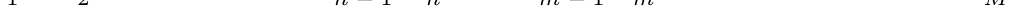}\\
\includegraphics[width=\graphiclength,clip]{ini.eps}
\caption{
Nanostructure attached to leads and 
schematic density profile of the initial wavepacket at $T=0$.
}\label{Fig:NanoSystem}
\end{center}
\end{figure}

The Hamiltonian for the nanostructure attached to leads, 
$\hat{H} = \hat{H}_{\mathrm{S}} \,+\, \hat{H}_{\mathrm{L}}
\,+\,\hat{H}_{\mathrm{C}}$ is given by
\begin{subequations} \label{eqn:Hamiltonian}
\begin{eqnarray}
\label{eqn:Hamiltonian-system}
\hat{H}_{\mathrm{S}} &=& 
  \sum_{j=n+1}^{m-1} -t_{\mathrm{S}} ( c_{j}^\dagger c_{j-1}^{\phantom\dagger} + \mathrm{H.c.} )
    + \sum_{j=n}^{m-1}\mu_g n_j
\nonumber\\ 
& & 
  +\, \sum_{j=n+1}^{m-1} V \left(n_{j}-\frac{1}{2}\right)\left(n_{j-1}-\frac{1}{2}\right), \\
\label{eqn:Hamiltonian-leads}
\hat{H}_{\mathrm{L}} &=& 
     \sum_{1 < j < n, m < j \leq M } -t ( c_{j}^\dagger c_{j-1}^{\phantom\dagger} +
                             \mathrm{H.c.}), \\
\label{eqn:Hamiltonian-contacts}
\hat{H}_{\mathrm{C}} &=& 
-t_{\mathrm{C}} (c_{n}^\dagger c_{n-1}^{\phantom\dagger} + 
     c_{m}^\dagger c_{m-1}^{\phantom\dagger} + \mathrm{H.c.}).
\end{eqnarray}
\end{subequations}
Individual sites are labeled according to Fig.~\ref{Fig:NanoSystem},
$M_S = m - n$ is the size of the interacting nanostructure, $\mu_g$
denotes a local external potential, which can be applied to the
nanostructure, and $V$ is a nearest-neighbor interaction term inside the
nanostructure. The hopping elements in the leads, the structure, and
coupling of the structure to the leads are $t$, $t_{\mathrm{S}}$, and
$t_{\mathrm{C}}$ respectively.
Similar to the approach in  \cite{peter04} we
add an external source-drain potential 
$\hat{H}_1 = \mu_{\mathrm SD}/2 \left(    \sum_{j=1  }^{n-1}\,  n_j \;
                               - \, \sum_{j=m}^{M}    n_j \right),$
to the unperturbed
Hamiltonian $\hat{H}$ and take the ground state $\ket{\Psi(T=0)}$ of
$\hat{H} + \hat{H}_1$, 
obtained by a standard finite lattice DMRG calculation, as initial
state at time $T=0$ \cite{peter04}. In addition we target for the ground state of $\hat{H}$.
In actual calculations the switched external potential was smeared out
over three lattice sites.

We then perform a time evolution as described above by applying the time evolution operator
$U={\mathrm e}^{i \hat{H} T}$ on $\ket{\Psi(T=0)}$ %
\footnote{It is equally possible to prepare the system in the groundstate of
the Hamiltonian $\hat{H}$, and to perform the time evolution via
$U={\mathrm e}^{i (\hat{H} + \hat{H}_1) T}$.}
\footnote{
We use a time step of 
$\Delta T = \hbar/2t$. In each adaptive time step one
intermediate time step is targeted and 2 sweeps are performed.},
which leads to flow of the extended wave packet through the whole
system until it is reflected at the hard wall boundaries as described
in \cite{peter04}.

The expectation value of the current at each bond and every time step
is given by
\begin{equation} \label{eqn:current-calc}
  J_{j,j-1}(t) = -\frac{2e}{\hbar} \mathrm{Re}\{
    {i}\langle\Psi(t)\vert 
      t_\mathrm{j}^{\phantom\dagger}
      c_\mathrm{j}^{\dagger}c_\mathrm{j-1}^{\phantom\dagger} 
    \vert\Psi(t)\rangle 
  \}.
\end{equation}
Following Refs.~\cite{wingreen93,cazalilla02} we define the current
through the nanostructure as an average over the current in the left and
right contacts to the nanostructure
\begin{equation} \label{eqn:current-average}
  J(T) = [ J_{n,n-1}(T) + J_{m,m-1}(T) ] / 2.
\end{equation}

For the calculation of the DC-conductance through the
nanostructure the 
time evolution has to be carried out for sufficiently long times
until a quasi-stationary state is reached and the steady state
current $J$ can be calculated.
If the stationary state corresponds to a well-defined applied external
potential $\mu_{\mathrm{SD}}$, the differential conductance is given by
$g(\mu_{\mathrm{SD}}) = e\, \partial J(\mu_{\mathrm{SD}}) / 
                                 \partial \mu_{\mathrm{SD}}.$ 
In the limit of a small applied potential, $\mu_{\mathrm{SD}}
\rightarrow 0$, the linear conductance is given by 
$g(\mu_{\mathrm{g}}) = e J(\mu_{\mathrm{g}}) / \mu_{\mathrm{SD}}.$

\begin{figure}[htb]
\begin{center}
\setlength{\graphiclength}{0.475\textwidth}
\includegraphics[width=\graphiclength,clip]{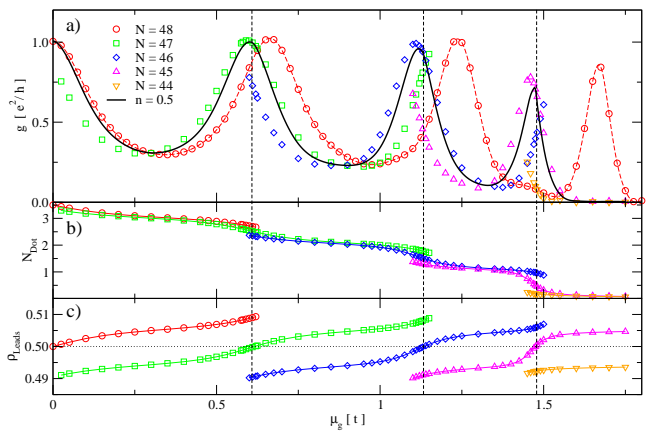}
\caption{
Transport through a non-interacting 7-site nanostructure with
$t_{\mathrm{C}}=0.5t$ and $t_{\mathrm{S}}=0.8t$. 
The energy levels of the nanostructure are indicated by dashed vertical lines.
(a) Linear conductance for different $N$.
The result after applying finite size corrections is shown as 
straight line (see text for details).
(b) Number of fermions on the 7-site nanostructure.
(c) Density $\rho=(N-N_{\mathrm{Dot}})/(M-M_s)$ in the leads.
System size is $M=96$ and the number of states kept in the DMRG is
$n_{\mathrm{Cut}}=400$.
}\label{Fig:gnf-d7v0t80t50}
\end{center}
\end{figure}

We first consider the transport through a single impurity.
\begin{figure}[tb]
\begin{center}
\setlength{\graphiclength}{0.475\textwidth}
\includegraphics[width=\graphiclength,clip]{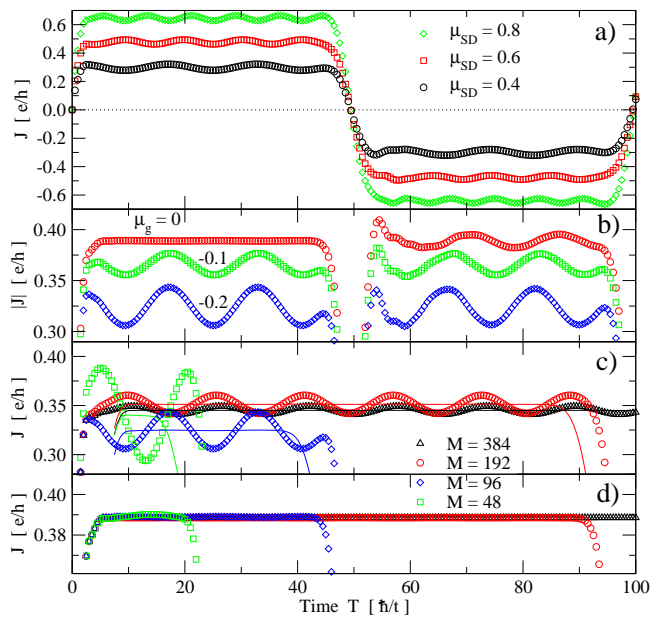}
\caption{
Current through a single impurity with $t_{\mathrm{C}}=0.5t$ at
nominal half filling $N/M = 0.5$ obtained from exact numerical
diagonalization.
(a) For different bias voltages ($\mu_g=-0.2$).
(b) For different gate voltages 
($\mu_{\mathrm{SD}}=0.4$).
System size is $M=96$.
(c,d) For different system sizes:
(c) $\mu_g=-0.2$, and
(d) $\mu_g=0$. (See text for details.)
}\label{Fig:tj-d1t50}

\end{center}
\end{figure}
The current rises from zero and settles into an oscillating quasi-stationary
state (Fig.~\ref{Fig:tj-d1t50}). After the wavepackets have travelled
to the boundaries of the
system and back to the nanostructure, the current falls back to zero
and changes sign. 
The amplitudes of the oscillations depend on $\mu_{\mathrm{SD}}$ and $\mu_g$,
and are proportional to the inverse of the system size $1/M$. 
The period of oscillation strongly depends on the applied potential
[Fig.~\ref{Fig:tj-d1t50} (a)] but is independent of
the gate potential and system size [Fig.~\ref{Fig:tj-d1t50} (b), (c)],
and is given by $T_{\mathrm{osc}}=2\pi\hbar/\vert \mu_{\mathrm{SD}} \vert)$.
This periodic contribution to the current is reminiscent of the Josephson 
contribution in the tunneling Hamiltonian, obtained by
gauge transforming the voltage into a time dependent coupling
$\tilde{t}_{\mathrm{C}}(T) = {t}_{\mathrm{C}} \,{\mathrm e}^{i \mu_{\mathrm SD}T}$.
It is present even 
for zero gate potential, but the currents in the
left and right leads oscillate with opposite phase and cancel in the
current average Eq.\,(\ref{eqn:current-average}).
After the wavepackets have finished one round trip, the current
oscillations reappear because of the
additional phase shift due to the different lengths of the left and
right leads [Fig.~\ref{Fig:tj-d1t50} (b)].
\begin{figure}[h]
\begin{center}
\setlength{\graphiclength}{0.475\textwidth}
\includegraphics[width=\graphiclength,clip]{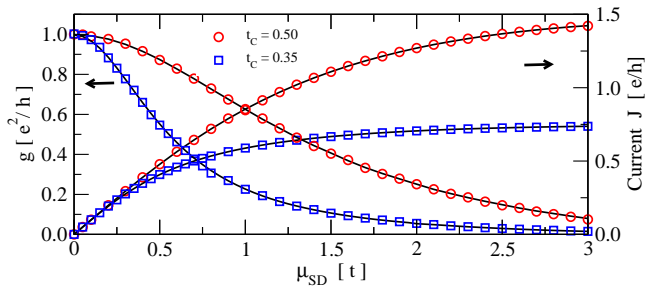}
\caption{
Current and differential conductance as function of applied potential
through a single impurity
with $\mu_g=0$ and half filled leads: $N/M = 0.5$. 
Circles (squares) show results for $t_{\mathrm{C}}=0.5t$ ($0.35t$). 
System size was $M=48$ ($M=96$) and $n_{\mathrm{Cut}}=200$ (400)
states were kept in the DMRG.
Lines are exact diagonalization results for $M=512$.
}\label{Fig:Jg-SingleImpurity}
\end{center}
\end{figure}
The stationary current is given by a straightforward average, 
because the oscillation period $T_{\mathrm{osc}}$ is known.
In general, the density in the leads, and therefore also the
current, depends on the system size and a finite size
analysis has to be carried out in order to extract quantitative results
[Fig.~\ref{Fig:tj-d1t50} (c), see also discussion of
Fig.~\ref{Fig:gnf-d7v0t80t50}].
Only in special cases (symmetry, half filled leads, and zero gate
potential) is the stationary current independent of the system size
[Fig.~\ref{Fig:tj-d1t50} (d)].
\begin{figure}[b]
\begin{center}
\setlength{\graphiclength}{0.475\textwidth}
\includegraphics[width=\graphiclength,clip]{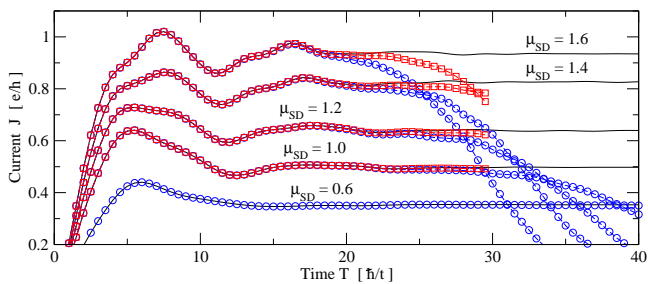}
\caption{
Exponentially growing error in the adaptive t-DMRG for large bias
voltages and times: Current through a non-interacting 7 site
nanostructure with $t_{\mathrm{C}}=0.5t$, $t_{\mathrm{S}}=0.8t$, and $\mu_g=0$. System size is
$M=144$ and $N/M= 0.5$.
The number of states kept in the DMRG were $n_{\mathrm{Cut}}=600$
(circles) and 1000 (squares).
Lines are exact diagonalization results. 
}\label{Fig:tj-d7v0t50}
\end{center}
\end{figure}

Our result for the conductance through a single impurity in
Fig.~\ref{Fig:Jg-SingleImpurity} is in excellent quantitative
agreement with exact diagonalization results already for moderate
system sizes and DMRG cutoffs. Accurate calculations for extended
systems with interactions are more difficult, mainly for two reasons: 
1.)~The numerical effort required for our approach depends crucially on
the time to reach a quasi-stationary state.
For the single impurity, the quasi-stationary state is reached on a
timescale proportional to the inverse of the width of the conductance
resonance, $4t\hbar/t_{\mathrm{C}}^2$, in agreement with the result in
Ref.~\cite{wingreen93}.
In general, extended structures with interactions will take longer to
reach a quasi-stationary state, and the time evolution has to be
carried out to correspondingly longer times. 
2.)~In the adaptive t-DMRG, the truncation error grows exponentially
due to the continued application
of the wave function projection, and causes the sudden onset of
an exponentially growing error in the calculated time evolution after
some time. This 'runaway' time is strongly dependent on the DMRG
cutoff, and was first observed in an adaptive t-DMRG study of spin
transport by Gobert et~al.\cite{gobert05}.
We observe the sudden onset of an exponentially growing error in our
calculations as well, Fig.~\ref{Fig:tj-d7v0t50}, but in addition to
the dependence on $n_{\mathrm{Cut}}$, the 'runaway' time now also
depends strongly on $\mu_{\mathrm{SD}}$. 
To avoid these problems one has to resort to the full t-DMRG
\cite{peter04}, which does not suffer from the runaway error.
A detailed analysis of the numerics of our approach will be published
elsewhere \footnote{G. Schneider and P. Schmitteckert, to be 
published.}.

\begin{figure}[tb]
\begin{center}
\setlength{\graphiclength}{0.475\textwidth}
\includegraphics[width=\graphiclength,clip]{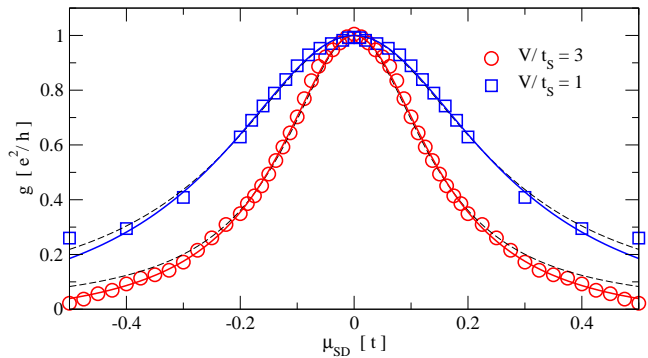}
\caption{
Differential conductance as a function of bias voltage
through a 7 site nanostructure with nearest neighbor interaction. 
Parameters are $t_{\mathrm{C}}=0.5t$, $t_{\mathrm{S}}=0.8t$, and N/M=0.5. 
Squares (circles) denote weak (strong) interaction with
$V/t_{\mathrm{S}}=1\;(3)$. Lines are fits to a Lorentzian with an energy
dependent self energy $\Sigma = {i}\eta_0 + {i}\eta_1 \mu^2$.
Dashed lines: $\eta_1=0$.
System size is $M=144$ ($M=192$) and 600 (800) states were kept
in the DMRG.
}\label{Fig:Jg-d7vX}
\end{center}
\end{figure}

In Fig.~\ref{Fig:Jg-d7vX} we show results for the first differential conductance peak
of an interacting 7-site nanostructure. Careful analysis of the data shows,
that in order to reproduce the  line shape accurately, one has to introduce an
energy dependent self energy. Since the effect is small,
we approximate it by a correction quadratic in the bias voltage
difference $\mu=\mu_\mathrm{SD}-\mu_\mathrm{peak}$.
It is important to note that for the strongly interacting nanostructure, $V/t_{\mathrm{S}}=3$,
the conductance peaks are very well separated. Therefore the line
shape is not overlapped by the neighboring peaks, and the fit is very
robust. Performing the same analysis for a non-interacting nanostructure
with a comparable resonance width, we obtain negligible
corrections to $\eta_1$ in the self energy,
indicating that the change of the line shape is due to correlation effects.

The linear conductance as a function of applied gate potential
can be calculated in the same manner, if
a sufficiently small applied external potential is used. 
We study the same non-interacting 7-site nanostructure as before and
use a bias voltage of $\mu_{\mathrm{SD}}=2\cdot10^{-4}$.
For half filled leads, the result for the linear conductance
calculated with a fixed number of fermions, $N/M=0.5$,
is qualitatively correct, but the conductance peaks are shifted to
higher energies relative to the expected peak positions 
at the energy levels of the non-interacting system
(Fig.~\ref{Fig:gnf-d7v0t80t50}). 
Varying the gate potential $\mu_g$ increases the charge on the nanostructure
by unity whenever an energy level of the nanostructure moves through the
Fermi level [Fig.~\ref{Fig:gnf-d7v0t80t50} (b)]. The density in
the leads varies accordingly 
[Fig.~\ref{Fig:gnf-d7v0t80t50} (c)]. Since the number
of fermions in the system is restricted to integer values, direct
calculation of the linear conductance at constant $\rho$ is not possible
and one must resort to interpolation.
Using linear interpolation in $\rho(N,\mu_g)$ for 
$N=44\dots48$ yields our final result for the linear conductance at
half filling [Fig.~\ref{Fig:gnf-d7v0t80t50} (a)]. The agreement in the
peak positions is well within the expected accuracy for a 96 site
calculation. 
\begin{figure}[b]
\begin{center}
\setlength{\graphiclength}{0.475\textwidth}
\includegraphics[width=\graphiclength,clip]{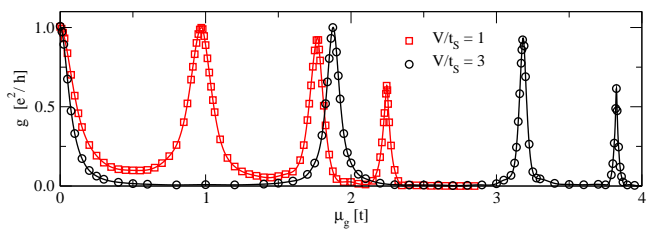}
\caption{
Linear conductance through an interacting 7 site system with
$t_{\mathrm{C}}=0.5t$ and $t_{\mathrm{S}}=0.8t$ for 
weak (squares) and
strong (circles) interaction. System size is $M=96$ ($M=192$) and
400 (600) states were kept in the DMRG.
Finite size corrections have been included.
Lines are guides to the eye.
}\label{Fig:gnf-d7vInt}
\end{center}
\end{figure}
Our results for the conductance through an interacting extended
nanostructure are presented in Fig.~\ref{Fig:gnf-d7vInt}. 
The calculation for the weakly interacting system requires roughly the same
numerical effort as the non-interacting system.
In the strongly interacting case, where the nanostructure is now in the
charge density wave regime, the time to reach a quasi-stationary state
is longer, and a correspondingly larger system size was used in the
calculation.
In both cases we obtain peak heights
for the central and first conductance resonance to within 1\% of the
conductance for a single channel.

We have introduced a new concept of extracting the finite bias and
linear conductance from real time evolution calculations. 
Very accurate
quantitative results are possible, as long as finite size effects are
taken into account.
Our results for the linear conductance compare favorably both in
accuracy and computational effort with the DMRG evaluation of the Kubo
formula \cite{dan06}.
Calculations of strongly interacting systems show correlation induced
corrections to the resonance line shape.

We profited from many discussions with Peter W{\"o}lfle.
The authors acknowledge the support from the DFG
through project B2.10 of the Center for Functional Nanostructures,
and from the Landesstiftung Baden-W\"{u}rttemberg under project B710.



\end{document}